\documentclass[letterpaper,twocolumn,10pt]{article}
\usepackage{usenix-2020-09}

\usepackage[english]{babel}
\usepackage{blindtext}
\usepackage{tikz}
\usepackage{amsmath}
\usepackage{enumitem}
\usepackage{caption}
\usepackage{subcaption}
\usepackage{xspace}
\usepackage{algorithm2e}
\usepackage{authblk}

\newcommand{\gossipsub}{{\texttt{GossipSub}}~}
\newcommand{\libpp}{{\texttt{libp2p}}~}
\newcommand{\icpp}{{\texttt{ICp2p}}~}
\newcommand{\todo}[1]{{\color{red}\emph{{[TODO]: #1}}}} 
\newcommand{\yap}[1]{{\color{magenta}\emph{{[YAP]: #1}}}} 
\newcommand{\tobias}[1]{{\color{purple}\emph{{[Tobias]: #1}}}} 
\newcommand{\yotam}[1]{{\color{blue}\emph{{[Yotam]: #1}}}} 
\newcommand{\ogi}[1]{{\color{olive}\emph{{[Ogi]: #1}}}} 
\newcommand{\rosti}[1]{{\color{green}\emph{{[Rosti]: #1}}}} 
\newcommand{\tim}[1]{{\color{orange}\emph{{[Tim]: #1}}}} 

\ifx\showcomments\undefined
    \renewcommand{\todo}[1]{} 
    \renewcommand{\yap}[1]{} 
    \renewcommand{\tobias}[1]{} 
    \renewcommand{\yotam}[1]{} 
    \renewcommand{\ogi}[1]{} 
    \renewcommand{\rosti}[1]{} 
    \renewcommand{\tim}[1]{} 
\fi

\newcommand{\ignore}[1]{}
\newcommand{\ic}{IC\xspace}
\newcommand{\icname}{\textit{Internet Computer}\xspace}

\begin{document}
\title{A New Broadcast Primitive for BFT Protocols}

\author{
Manu Drijvers$^1$,
Tim Gretler\thanks{Work done while at DFINITY Foundation}~,
Yotam Harchol$^{3*}$,
Tobias Klenze$^*$,
Ognjen Maric$^1$,
Stefan Neamtu$^*$,
Yvonne-Anne Pignolet$^1$,
Rostislav Rumenov$^1$,
Daniel Sharifi$^1$,
Victor Shoup$^{3*}$\\

\large{$^1$DFINITY Foundation, $^2$NVIDIA, $^3$Offchain Labs}

\small{\texttt{manu@dfinity.org},
       \texttt{gretler.tim@gmail.com},
       \texttt{yotam@cs.berkeley.edu},
       \texttt{tobias.klenze@stusta.net},
       \texttt{ognjen.maric@dfinity.org},
       \texttt{stefan.neamtu@gmail.com},
       \texttt{yvonneanne@dfinity.org},
       \texttt{rostislav.rumenov@dfinity.org},
       \texttt{daniel.sharifi@dfinity.org},
       \texttt{victor@shoup.net},
       }
}

\date{}

\maketitle

\begin{abstract}
Byzantine fault tolerant (BFT) protocol descriptions often assume application-layer networking primitives, such as best-effort and reliable broadcast, which are impossible to implement in practice in a Byzantine environment as they require either unbounded buffering of messages or giving up liveness, under certain circumstances.
However, many of these protocols do not (or can be modified to not) need such strong networking primitives. 
In this paper, we define a new, slightly weaker networking primitive that we call \emph{abortable broadcast}. We describe an implementation of this new primitive and show that it (1) still provides strong delivery guarantees, even in the case of network congestion, link or peer failure, and backpressure, (2) preserves bandwidth, and (3) enforces all data structures to be bounded even in the presence of malicious peers. The latter prevents out-of-memory DoS attacks by malicious peers, an issue often overlooked in the literature.
The new primitive and its implementation are not just theoretical. We use them to implement the BFT protocols in the \icname (\ic), a publicly available blockchain network that enables replicated execution of general-purpose computation, serving hundreds of thousands of applications and their users.
%
%
In addition, we provide evaluation results based on experimental data and measurements of the implementation running on the \ic mainnet.

\end{abstract}


\section{Introduction}

With the advent of blockchain technology, Byzantine fault-tolerant (BFT) protocols have left the realm of pure academia. They are now securing daily transactions worth billions of USD for millions of users worldwide. Many BFT protocols, from consensus protocols\cite{pbft,buchman_tendermint_2016,buterin_casper_2019}, threshold signature schemes\cite{aumasson_survey_2020}, to multi-party computation protocols\cite{lu_honeybadgermpc_2019,das_practical_2022} rely on the humble \emph{broadcast} networking primitive as a basic building block. 
In this paper, (i) we show that this trivial-looking primitive is challenging to implement in a Byzantine environment in a way that preserves liveness, and (ii) we offer a general-purpose solution to these challenges by proposing a new broadcast primitive useful for a wide class of BFT protocols.

Broadcast allows a sender node to send a message to all the members of some group. While this primitive is familiar from lower levels of the communication stack (e.g., OSI layers 2 and 3), in this paper we exclusively focus on a different broadcast primitive that is used between an overlay network of peers and implemented on the application layer (Layer 7). 
The distributed systems literature that deals with BFT protocols formally defines many different flavors of broadcast\cite{cachin2011introduction}, but generally requires broadcast to at least (1) be non-blocking (in the sense that invoking the primitive will return control in bounded time), and  (2) provide guaranteed delivery in the following sense: \emph{all} messages that are broadcast from a correct (non-faulty) sender are guaranteed to get (eventually) delivered to all correct recipients. However, these simple requirements turn out not to be implementable in practice when combined with a third requirement that's often neglected in the literature, but is no less important in practice: bounded memory usage.

\begin{figure*}[t]
\captionsetup[subfigure]{labelformat=empty}
    \begin{subfigure}{.5\linewidth}
        \includegraphics[width=0.9\textwidth]{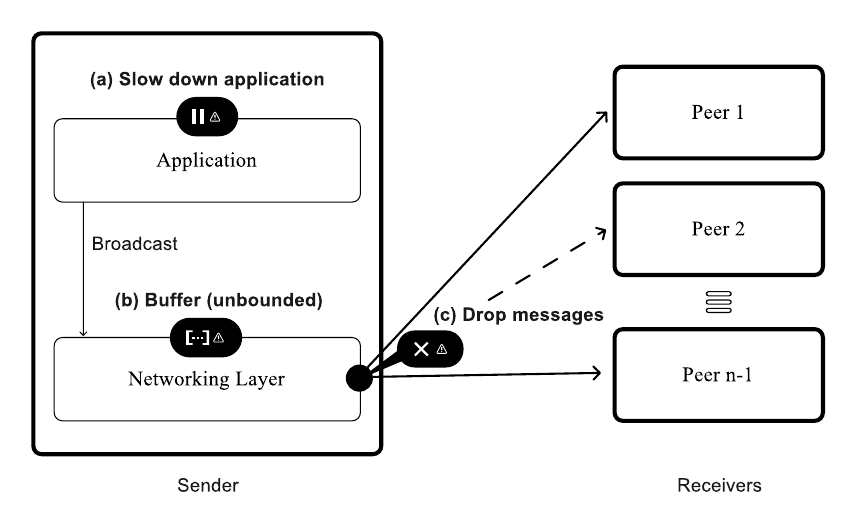}  
    \vspace{-.2cm}
       \caption{Traditional broadcast}
    \end{subfigure}
    \begin{subfigure}{.5\linewidth}
        \includegraphics[width=0.9\textwidth]{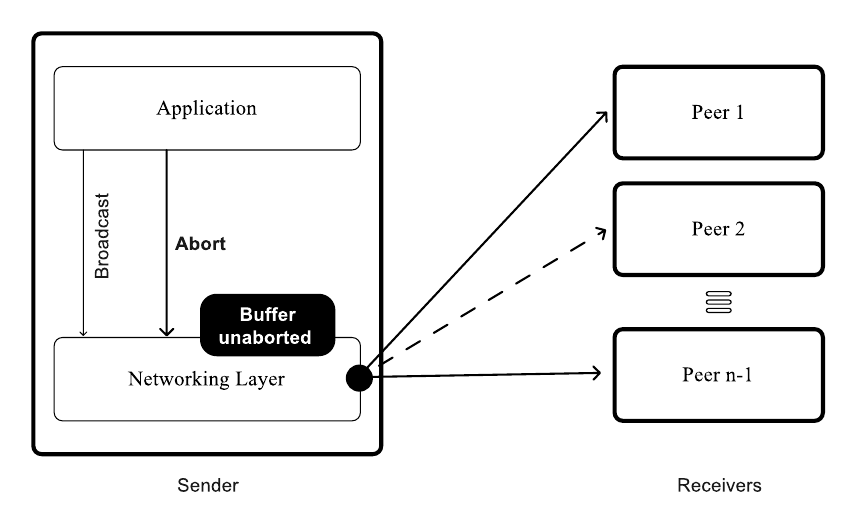}
    \vspace{-.2cm}
        \caption{Abortable broadcast}
    \end{subfigure}
    \vspace{-.4cm}
    \caption{\emph{(left)} Traditional broadcast protocols' core challenge is reacting to backpressure, either due to slow or faulty peers or links or malicious peers' behavior. Slowing down the application, buffering indefinitely, and dropping messages are all unacceptable. \emph{(right)} Abortable broadcast overcomes these drawbacks and can be implemented efficiently with bounds on memory and delivery guarantees.}
    \label{fig:backpressure}
    \vspace{-.4cm}
\end{figure*}

The core challenge, shown in Figure~\ref{fig:backpressure}\emph{(left)}, is reacting to backpressure due to network or peer problems. To understand this better, let's first consider a traditional point-to-point networking layer, such as the one used between a Web client and a server. Here, and in the rest of the paper, we use the term \emph{networking layer} in reference to user-space code, possibly running on top of the TCP, UDP, or QUIC stacks, that provides network access to other parts of the application (clients). Such a networking layer generally has three options to handle backpressure. On the send side, it can (a) propagate the backpressure to the client (e.g., by blocking the client), and hence slow down data production, (b) buffer an unbounded number of egress messages, or (c) drop egress messages. For example, a networking layer implemented over TCP can use options (a) or (b).  When the receiving end of the TCP channel is slower than the sending end,  the TCP protocol temporarily stores data in a buffer. Once the buffer is filled, the caller (networking layer) is backpressured. The networking layer can choose option (a) to simply propagate the backpressure to their clients, which can then handle it, generally by slowing data production down. This makes perfect sense in this setting: as long as the (sole) recipient cannot receive the data, there is no point in producing more of it.

But for broadcast as used by BFT protocols, none of these options work. (a) means that one slow peer, which can potentially behave so on purpose, can slow down or even halt the BFT protocol. (b) means that the slow peer can push other peers to their memory limits, crashing the peers and thus causing a DoS. (c) means that if a link is suffering from actual congestion, or a benign peer is honestly slow, the networking layer would violate the delivery guarantee, creating a risk of that peer getting stuck forever by missing an important message, which can, in turn, destroy the liveness guarantees of the entire protocol. Of course, the last risk can be mitigated with retransmission at the networking layer level, but the retransmission itself is susceptible to the same problems.

Thus, our three requirements, non-blocking broadcast, guaranteed delivery, and bounded memory usage seem impossible to reconcile. But we show that most practical protocols don't need to reconcile them. Namely, the guaranteed delivery property is normally used in theoretical high-level descriptions of BFT protocols and their correctness proofs to argue that a peer who may be arbitrarily far behind can still catch up.
However, practical implementations of these protocols generally perform some modifications to such high-level descriptions of protocols. In particular, they use \emph{checkpointing} to regularly summarize the current protocol state, making all previous messages obsolete, and thus bounding the number of relevant messages at any given point in time\cite{pbft,stathakopoulou_mir-bft_2019,avarikioti_fnf-bft_2023}.
To ensure bounded memory usage, these practical protocol implementations also require a practical counterpart to the reliable broadcast primitive that can take advantage of messages becoming obsolete.
It is exactly this piece that is usually overlooked in the literature and that we provide in this paper. Specifically, our four core contributions are as follows.

    \textbf{Abortable broadcast} We define a generic \emph{abortable broadcast} interface that is non-blocking, but slightly weakens the promise of guaranteed delivery to \emph{abortable guaranteed delivery}. This weakening is key to enabling memory-bounded implementations, and it requires clients to explicitly abort broadcasting obsolete messages, illustrated in Fig.~\ref{fig:backpressure}\emph{(right)}.
    Protocol messages can become obsolete as protocols are progressing; clients only need guaranteed delivery of non-obsolete, \emph{active} messages. 
    As long as the number of \emph{active} (non-aborted) messages is bounded, they are guaranteed to be delivered, and within bounded time at that whenever the network is well-behaved. This holds regardless of the rate of message creation and deletion (which can be very high) or the rate of data transmission to any specific peer (which can be arbitrarily low).    
    This weaker interface still suffices to implement practical client protocols. In particular, we show that we can use this new primitive to implement all the BFT protocols used in the \icname (\ic) blockchain as clients of our abortable broadcast implementation, such as the core consensus protocol \cite{camenisch2022internet}, multiple threshold signature protocols \cite{tecdsa}, and distributed key generation \cite{dkg}.
    The interface is applicable beyond just the \ic protocols: it is likely sufficient for any BFT protocol that involves a bounded number of peers and uses checkpoints.

     \textbf{Slot table} We show how to implement abortable broadcast with a novel P2P protocol using strictly bounded data structures and authenticated communication channels. We define a simple data structure that we call a \emph{slot table}. Given a maximal number of active messages that a client requires for its successful operation at any point in time, the slot table tracks the creation and deletion of messages using bounded memory, and the P2P protocol distributes and disseminates them in the network. The speed of sending updates to the slot table is adjusted for each peer separately, without propagating the backpressure to the client and without slowing down message distribution to other peers. 
    To the best of our knowledge, this is the first published peer-to-peer protocol that provides delivery guarantees with bounded data structures in a Byzantine setting of a geographically dispersed peer-to-peer network.

    \textbf{Complete and efficient implementation} The networking layer described in this paper is fully implemented (including optimizations to meet performance requirements) and has been running in production on 560 nodes, spread across the globe, for over a year.  
    We also discuss important implementation design decisions as we believe this can be useful for others who build similar systems.

    \textbf{Evaluation} 
        We evaluate our approach with a security analysis and by comparing its performance to that of \gossipsub \cite{gossipsub}, a state-of-the-art P2P library that is part of \libpp \cite{libp2p}, used in many blockchain projects. \gossipsub is designed to work with larger networks than our protocol, but unlike our protocol, it does not provide any delivery guarantee. In small and medium network sizes, such as used by the \ic, our implementation's performance compares favorably. 

    

Finally, we end the paper with a call for collaboration with the academic community and a list of possible future work in the networking domain. We believe that blockchain-based software execution offers significant advantages over legacy cloud computing for some applications, and that networking is a foundational aspect in blockchain protocol design and implementations. This emerging technology still has much room for improvement and research. 

\textit{Ethics statement.} We disclaim ethical issues in this paper.

\section{Background}
\label{sec:background}

\label{sec:background:ic}
Modern blockchain networks aim at providing efficient multi-tenant, general-purpose, and tamperproof computation, in a decentralized and geo-replicated manner. To achieve this, these networks rely on multiple distributed protocols, most of which can be categorized as BFT protocols.

\paragraph{BFT Protocols}
BFT protocols are designed to operate correctly even if some of the participants (called \emph{nodes} or \emph{peers}) try to undermine it. We denote nodes that adhere to the protocol and are up and running as \emph{correct} or \emph{honest}, and all others as \emph{faulty} or \emph{malicious}. In section~\ref{sec:design}, we will relax this definition to not consider nodes that crash and recover faulty. 

A prime example of such protocols is the family of \emph{consensus} protocols,  which blockchains use as the basis for state machine replication. A blockchain consensus protocol produces a series of \emph{blocks}, with each block containing a series of decisions. These decisions can be financial transactions, or, as in the \ic, inputs for software running as replicated state machines. Denoting the number of faulty nodes by $f$, and the total number of nodes by $n$, asynchronous and partially-synchronous consensus protocols require that $f < n / 3$~\cite{lamport_byzantine_1982}.

Any BFT protocol maintains a pool of messages that need to be propagated to other peers that participate in the protocol. The pool is populated with incoming messages as well as with messages produced locally by the protocol, which often need to be broadcast to all other peers. Messages can also be removed from the pool when they are no longer needed.

The message pools of many BFT protocols, including the ones used by the \ic, share two important properties that will enable important design decisions we explain later:
\begin{enumerate}[label=(P\arabic*)]
    \item \textbf{Explicit Deletion of Messages.} Messages deleted from the pool no longer need to be disseminated to peers. Or, from the receiver side, only messages that at least one other peer has in their pool may need to be obtained. Practical implementations of BFT protocols satisfy this property as they use checkpoints, allowing them to purge messages periodically.  \label{prop:expiry} 
    \item \textbf{Bounded Active Messages.} At any point in time, the number of messages in the pool and their size is bounded. The maximal number of artifacts $C$ in the pool is a function of the checkpoint interval (for example, in consensus protocols this can be measured in the number of blocks or rounds), and the number of peers. The size of each message is bounded as well.  \label{prop:bounded}
\end{enumerate}

\paragraph{Traditional Broadcast Properties}

The distributed computing literature defines many variants of broadcast algorithms. The broadcast interface enables the client on the sender side $s$ to invoke a \emph{broadcast($m$)} operation for a message $m$, which eventually triggers a \emph{deliver($s$, $m$)} event on the receiver side client. Some broadcast variants change the receiver interface to trigger \emph{deliver($s$, $l$, $m$)} events, where $l$ is some label generated by the broadcast algorithm. The different broadcast variants then provide different properties under different failure models. Common properties (taken from~\cite{cachin2011introduction}) include:
\begin{itemize}
    \item \textbf{Validity}: If a correct node broadcasts a message $m$, then every correct node eventually delivers $m$.
    \item \textbf{Integrity}: If a node delivers a message $m$ from a correct sender $s$, then $s$ has previously broadcast $m$ (i.e., no tampering happened).
    \item \textbf{Consistency}: (Needs a label $l$). Whenever correct nodes deliver messages $m$ and $m'$ with the same sender $s$ and label $l$, then $m = m'$. 
    \item \textbf{Totality}: (Needs a label $l$). If any correct node delivers a message from a sender $s$ with label $l$ then all correct nodes eventually deliver a message from $s$ and label $l$.
\end{itemize}
For example, BFT reliable broadcast guarantees all these properties, while consistent broadcast does not guarantee the totality property, and best-effort broadcast (somewhat confusingly named, as it still guarantees delivery for correct senders) only guarantees the first two properties~\cite{cachin2011introduction}.\footnote{Best-effort broadcast is not defined for the BFT setting in~\cite{cachin2011introduction}, but the definition for the benign failure setting transfers to the BFT one. We also omit the ``no duplication'' property from~\cite{cachin2011introduction} above, as it is not so useful in the BFT setting.} 

The \emph{totality} and \emph{consistency} properties can be used to prevent \emph{partition attacks}, where a malicious node sends one message to a part of the peers, and a contradicting message to another part of the peers, for instance in equivocation attacks in consensus. 
Achieving these properties requires adding more traffic, for example, relaying the received messages or their error-coding fragments to all other peers. In turn, this increases both the latency (as multiple message delays are now needed) and message complexity. For this reason, most higher-level BFT protocols (including all the ones presented in~\cite{cachin2011introduction} and the ones used in the \ic) rely on best-effort broadcast instead, and relay messages themselves. This way, they can pick which messages to relay, preserving bandwidth. Even for the messages that they do relay, they can use their knowledge of the message content and state to be more efficient compared the the generic reliable broadcast primitive. We thus focus on the validity and integrity properties in the rest of the paper, and we'll omit the label $l$ from the interface, since it is not used by the properties we care about. 

\section{Abortable Broadcast}
\label{sec:abortable-broadcast}

Broadcast variations from Section~\ref{sec:background} are taken as basic modular building blocks and assumed straightforward to implement. However, there are several issues that hinder implementing them in practice, which we discuss in Section~\ref{sec:ab:rb-problems}. We alleviate the most important problem with a new broadcast variant, \emph{abortable broadcast}, in Section~\ref{sec:ab:definition}. In Section~\ref{sec:design} we will then define an optimized practical version of the abortable broadcast interface that we use in the \ic. 

\subsection{Issues with existing broadcast definitions}
\label{sec:ab:rb-problems}

The crucial problem with the broadcast definitions from Section~\ref{sec:background} is that the basic property, \emph{validity}, which describes guaranteed eventual delivery, is problematic to implement in practice. Practical protocols run continuously, and continuously produce messages. Reliably delivering all of them can require unbounded message storage for retransmission. This may create a denial-of-service (DoS) vulnerability, where malicious nodes may cause others to buffer messages until they go out of memory, crash, or just be very slow, for example, due to virtual memory thrashing. Of course, messages can be stored in persistent storage instead of RAM, but this makes the implementation significantly more complicated and less performant, and doesn't solve the conceptual problem, as the persistent storage will also be exhausted at some point.

This problem is a serious concern in practice. As we show in the evaluation section, it causes some existing practical broadcast solutions (in particular~\cite{libp2p}) to simply give up the validity property, knowing it cannot be implemented in practice without creating security vulnerabilities. However, by doing so, they burden their clients with implementing fallbacks for failed message deliveries. Clients that do not handle this properly risk halting under unfavorable network conditions. Yet, this is seen as preferable compared to the alternative of memory exhaustion.

The second problem is that the broadcast interface defined in Section~\ref{sec:background} is suboptimal with respect to bandwidth usage. Even without checkpointing, not all undelivered message are useful for the receiver at the moment. For example, some might come from senders that are farther behind in the protocol and cannot help the receiver advance their state.

The last problem is that the definition of faulty nodes from Section~\ref{sec:background} includes nodes that crash and recover, which simplifies reasoning about (theoretical) BFT protocols. But in practice, since many BFT protocols such as blockchains are designed to run forever, this leads to an ever-increasing fraction of faulty nodes. Practical protocols thus typically need to account for this and allow a crashed node to recover.

We address the first problem now, and the other two in Section~\ref{sec:design:interface}.

\subsection{Abortable Broadcast Definition}
\label{sec:ab:definition}

We now define a new flavor of broadcast that does not suffer from the issue of unbounded memory usage, and  still enables straightforwardly implementing many modern BFT protocols used in practice, without sacrificing liveness in corner cases. 

Our key observation is that the validity property, which causes unbounded memory usage, is not needed for BFT protocols that satisfy the property~\ref{prop:expiry} from Section~\ref{sec:background}. For such protocols, old protocol messages get superseded (usually by checkpoints) and no longer need to be delivered. 


We use this observation to formally define the \textbf{\emph{abortable broadcast}} primitive. Its interface extends that of traditional broadcast primitives from Section~\ref{sec:background} with an additional \emph{abort($m$)} operation. It allows a sender to abort a message if they think that it is not needed by other nodes anymore. Then, abortable broadcast provides the following properties:

\begin{enumerate}[label=(AB\arabic*), left=0pt]
    \item \textbf{Abortable validity}: If a correct node sends a message $m$, then every correct node eventually delivers $m$, \emph{unless the sender aborts $m$}.\label{def:abortable-validity}
    
    
    \item \textbf{Integrity}: If a node delivers a message $m$ from a correct sender $s$, then $s$ has previously broadcast $m$ (i.e., no tampering happened).\label{def:integrity}
\end{enumerate}

Note that this is almost identical to the best-effort broadcast variant from Section~\ref{sec:background}, just with a weaker validity property. 

As noted in Section~\ref{sec:background}, many BFT protocols (typically those using checkpoints and having a bounded number of peers) satisfy~\ref{prop:bounded}: there is a known bound $C$ on the number of messages needed to be broadcast at any point in time. That is, the number of sent, but not aborted messages from any sender can always be kept below $C$. In this case, and assuming there is an upper bound on the message size, the abortable broadcast primitive must additionally satisfy the following property:


\refstepcounter{enumi}
\begin{enumerate}[label=(AB\arabic*), left=0pt, resume]
    \item \textbf{Memory boundedness} The total memory used by an abortable broadcast instance is below some bound $B$.\label{def:memory-boundedness}
\end{enumerate}

\subsection{The Slot Table Algorithm}
\label{sec:ab:algo}

\begin{algorithm}[!t]
\DontPrintSemicolon
\SetAlgoNoEnd
\SetAlgoLined
\SetKwFunction{Bcast}{broadcast}
\SetKwFunction{OnRecv}{on\_receive}
\SetKwFunction{Retransmit}{retransmit}
\SetKwFunction{Abort}{abort}
\SetKwFunction{Deliver}{deliver}
\SetKwFunction{Send}{send}
\SetKwData{None}{none}

\KwData{$C$: capacity;
\\~~$V \gets 0$: global version;
\\~~$SS[1..C]$ : send slots with fields $version$ and $msg$
\\~~$RS_s[1..C]$: receive slots with fields $version$ and $msg$}

\BlankLine

\SetKwProg{Fn}{Function}{}{}
\Fn{\Bcast{$m$}}{
    $i \gets \min \{ k \in [1,C] \mid SS[k].msg = \None \}$\;
    $V \gets V + 1$\;
    $SS[i] \gets \{ version = V, msg = m \}$\;
    \ForEach{peer $s$}{
        \Send($s$, $i$, $V$, $m$)
    }
}

\vspace{1em}

\SetKwProg{Fn}{Function}{}{}
\Fn{\Abort{$m$}}{
    $i \gets \min \{ k \in [1,C] \mid SS[k].msg = m \}$\;
    $V \gets V + 1$\;
    $SS[i] \gets \{ version = V, msg = \None \}$\;
}
\vspace{1em}

\SetKwProg{Fn}{Function}{}{}
\Fn{\OnRecv{$s$, $slot$, $v$, $m$}}{
    \If{$v > RS_s[slot].version$}{
        $RS_s[slot] \gets \{ version = v, msg = m \}$\;
        \Deliver{$s$, $m$}\;
    }
}

\vspace{1em}

\SetKwProg{Fn}{Periodically}{}{}
\Fn{\Retransmit{}}{
    \ForEach{peer $s$}{
        \For{$i \gets 1$ \KwTo $C$}{
            \If{$SS[i].msg \neq \None$}{
                \Send{$s$, $i$, $SS[i].version$, $SS[i].msg$}\;
            }
        }
    }
}

\vspace{1em}

\caption{Slot Table Synchronization Algorithm using authenticated
  unreliable channels for communication.}\label{alg:slot-table}
\end{algorithm}

Algorithm~\ref{alg:slot-table} shows a simple algorithm implementing the
abortable broadcast primitive. It illustrates the basic ideas behind our
approach; we will further develop it into a fully practical implementation later.

The algorithm is based on the \emph{slot table} data structure.  A
slot table is a numbered array of slots of capacity $C$, where $C$ is the
bound on the number of active messages assumed in Section~\ref{sec:ab:definition}. Each slot has a version number and is marked
either as free or contains a message. For simplicity, we conceptually assume
unbounded version numbers; in practice, using 64-bit numbers suffices to avoid
rolling over. Each node maintains a single \emph{send side slot table} $SS$, and
$n-1$ \emph{receive side slot tables} $RS_s$, one for each peer $s$.

When the client triggers a broadcast, the algorithm finds a free slot in the
send side table (guaranteed to exist, by the condition on $C$), increments the
version, and writes the message with the version in the free slot. Then it sends
a single message to all peers on an authenticated unreliable channel (e.g., a
signed message over UDP), announcing a new message in the slot where the version
has been updated. When the client aborts a message, the corresponding slot is
marked as free. Abortions do not have to be explicitly announced, as they will
eventually be overwritten by new messages. When a message is received from a
peer, the receiver compares the received version with the existing version in
the specified slot, and, if newer, delivers the message to the client.

\begin{figure}[t]
    \centering
    \includegraphics[width=0.78\columnwidth]{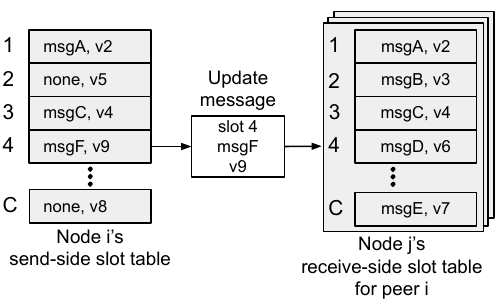}\vspace{-.4cm}
    \caption{Slot table data structure. Peers synchronize the view of each other's message pool in an eventually consistent protocol, with strict bounds.}
    \label{fig:slot-table}
\end{figure}

Figure~\ref{fig:slot-table} depicts this process with an example: Messages $A$ through $E$ were delivered from node $i$ to node $j$. Message $D$ (slot 4, version 6) as well as messages $B$ and $E$ were deleted, and message $F$ was created. Message $F$ was placed in slot number 4, and so an update message with the new message and version number 9 is sent to node $j$ (and all other peers).

Of course, since the messages are sent over an unreliable channel, they may get lost. The sender thus periodically retransmits all messages in
its send side slot table. The crucial observation is that the slot table is
bounded to at most $C$ slots. Thus, not all old messages need to be
retransmitted; only the latest version of each slot is retransmitted. This
still suffices to achieve the abortable guaranteed delivery property~\ref{def:abortable-validity}, while keeping the memory
usage of the algorithm bounded (achieving~\ref{def:memory-boundedness}).
Finally, the integrity property~\ref{def:integrity} is achieved thanks to the
authenticated channel.

\section{Design}
\label{sec:design}
In this section, we propose a design for a networking layer for the \ic protocols that satisfy Properties \ref{prop:expiry} and \ref{prop:bounded}. 
This networking layer implements the abortable broadcast primitive, and uses the
basic ideas presented in Algorithm~\ref{alg:slot-table}. However, it makes the
algorithm more practical in two ways: it (1) optimizes the bandwidth usage,
(2) uses a connection-based transport protocol instead of an unreliable channel,
and (3) supports node recovery, i.e., a node that crashes but then recovers does not count as faulty.

\begin{figure}[t]
    \centering
    \includegraphics[width=0.36\textwidth]{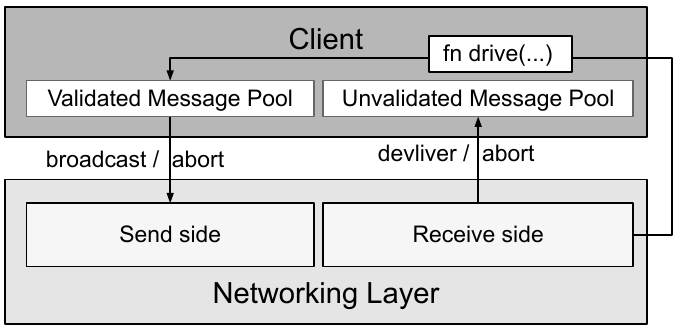}
    \caption{High-level interface of the networking layer with its clients.}
    \label{fig:interface}
    \vspace{-.4cm}
\end{figure}


\subsection{Client Interface and Guarantees}
\label{sec:design:interface}
As described in Section~\ref{sec:background}, the BFT protocols running on top of the networking layer presented in this section, collectively referred to as \emph{clients}, use a \emph{message pool} each to store and maintain the set of messages they currently need. These messages correspond to the current \emph{state} of the client. The messages from this pool are to be broadcast to all other peers. Our interface directly observes additions and deletions to the pool on the send side and triggers the corresponding nonblocking IO operations - \emph{broadcast} and \emph{abort}.

On the receiver side, our networking layer writes the broadcast messages to the
\emph{unvalidated} message pool to \emph{deliver} them. The ``unvalidated'' part
of the name reflects the fact that, since peers are untrusted, the client must
validate any message coming from a peer (or just discard it, for example if
it's obsolete). Moreover, our networking layer does not make any assumptions on the ordering of sent messages, nor does it provide any ordering guarantees on message delivery. Thus, if validating a received message requires some other messages as context, it must wait until these messages are also received. Therefore, the receive side also uses a pool of messages. While we believe that our approach could be extended to also provide ordering guarantees on message delivery, taking advantage of that would have required significant changes to the existing clients, so we did not pursue this further in this work. Finally, we note that we use a single, joint unvalidated pool for all senders, as this allows us to not deliver the same message more than once if it is sent by multiple senders.

Once the client validates a message from the unvalidated pool, based on its local information (e.g., checking signatures and block hashes), it can move the message to the send pool, which, for this reason, we call the \emph{validated} pool. In general, the client does not add messages to the unvalidated pool on its own, but can otherwise modify the pools as it wishes. It can remove messages, move them from unvalidated to validated, or even produce new messages based on the pools content or external data and put them directly into the validated message pool, which would lead to the broadcasting of these new messages. But it must keep the validated pool bounded; the networking layer requires each client to maintain an explicit bound $C$ on the size of this pool (recall Property \ref{prop:bounded} required from the BFT protocols). In turn, the networking layer keeps the unvalidated message pool bounded in size, such that no single peer, or a subset of them, could fill up the pool and prevent the receiver from receiving messages from other peers.

Figure~\ref{fig:interface} depicts the interface between the networking layer and its clients. All invocations are initiated by the networking layer, though a different design could also be implemented instead. The client interface is designed as a state machine that operates on the message pools. It is triggered by the networking layer via calls to the public \texttt{drive()} function. This function is called whenever there is an update to the unvalidated message pool, or when a timeout expires without updates (to let the client produce new messages if needed).

The \texttt{drive()} function returns a list of actions performed on the
validated message pool. These can be of two types: additions and deletions of
messages. The networking layer uses these actions to track the contents of the
validated pool: addition means a new message is to be broadcasted to all peers,
and deletion means aborting the broadcast of an existing message. The validated pool is persisted on disk, surviving crashes.
 
\paragraph{Bandwidth Preservation}
The traditional broadcast interface doesn't allow the receiver to control which messages it receives, potentially wasting bandwidth by downloading unneeded messages. To counter this, our clients provide a \emph{bouncer function} to determine if a message is needed. On the implementation side, we use \emph{adverts}: senders advertise available messages to their peers using small advert messages, instead of sending the messages directly, and receivers only download messages accepted by the bouncer function. An added benefit of adverts (even without using the bouncer function) is that if the same message is sent by multiple senders (e.g., for client-level relays to simulate reliable broadcast), it only needs to be downloaded once.

However, adverts introduce both latency and bandwidth overheads. They are therefore only used when the message size exceeds a certain threshold. In our implementation, we set this threshold to $1\textit{KB}$. We also provide the option for the client to specify that a certain message must be sent directly (that is, without an advert) so that the client can choose to push certain messages faster to peers.



\paragraph{Guarantees}
The goal of the networking layer is to disseminate messages in the validated pool of senders in the peer group through abortable broadcast. This guarantees delivery (addition to the unvalidated pool of honest receivers) for messages unless they are aborted through deletion from the validated pool of the senders or not requested by the receiver's bouncer function.

More formally, assuming that the client ensures that the validated pool of every correct node always contains no more than $C$ messages, and given a sender client $S$ and any receiver client $R$, both on correct nodes from the same peer group, our interface guarantees the following:
\begin{enumerate}[label=(G\arabic*)]
    \item \textbf{Abortable guaranteed delivery}: Whenever $S$ adds a message $m$ to
      its validated pool, if $S$ doesn't eventually delete $m$, and if $R$ is
      eventually forever willing to download $m$ (as indicated by the bouncer
      function), then $m$ is eventually delivered to $R$'s unvalidated
      pool. \label{guarantee:delivery}
    \item \textbf{Integrity}: If a message $m$ is delivered to $R$'s unvalidated pool with a correct sender $S$, then $S$ has previously broadcast $m$ (i.e., no tampering happened).\label{guarantee:integrity}
    \item \textbf{Bounded data structures}: The unvalidated pool, as well as all of the networking layer internal data structures, are bounded in size. \label{guarantee:bounded-ds}
    \item \textbf{Bounded-time delivery}: There's a time limit $\Delta$ such
      that in any $\Delta$-long period where the link between $S$ and $R$ is
      behaving correctly (i.e., $S$ and $R$ have their fair share of the claimed
      throughput/latency properties), whenever $S$ adds a message $m$ before the
      start of the period, doesn't delete it by the end of the period,
    and the bouncer function at $R$ constantly accepts $m$, then $m$ is delivered to $R$'s unvalidated pool in that period. \label{guarantee:partial-sync}
\end{enumerate}

\ref{guarantee:delivery}--\ref{guarantee:bounded-ds} are straightforward adaptations of the Properties~\ref{def:abortable-validity}--\ref{def:memory-boundedness} from Section~\ref{sec:abortable-broadcast}.
Property~\ref{guarantee:partial-sync} is required for liveness proofs of BFT
protocols in the partial synchrony model. It does not appear in the definitions
of Section~\ref{sec:background}, as those assume an asynchronous model
(potentially expanded with failure detectors).
As we will see later, our implementation ensures these properties hold even when correct nodes can crash and recover, as long as they are up-and-running for long enough, making the assumptions from Section~\ref{sec:background} more practical.

\subsection{Connection-oriented transport}
\label{sec:design:transport}

Algorithm~\ref{sec:ab:algo} used unreliable channels such as UDP to focus on the main conceptual points. However, UDP has a number of drawbacks that make it unsuitable for the networking layer in a practical setting: limited message length, no handling of packet loss, and no congestion control.

This is why in our implementation we rely on a connection-oriented transport, in particular QUIC with TLS instead. In Section~\ref{sec:impl}  we explain the choice of QUIC in more detail and how we implement \emph{broadcast} and \emph{abort} to be nonblocking.  

\subsection{Bootstrapping and Recovery}
\label{sec:design:recovery}

Membership in a peer group may change over time. We assume that there exists some external mechanism for managing the membership. For example, in the \ic, the group membership is managed by a decentralized voting procedure. What matters from a networking perspective is that existing nodes in the group should connect to newly joined nodes, and newly joined nodes should connect to other nodes in the group. If a node was removed, the connections to it should be dropped and no further connections from it should be accepted.

The process of a node joining a group is actually the same as that of a node
recovering from a crash since the slot table isn't persisted and is lost on
crashes. In both cases the node will start with an empty slot table.
Furthermore, when a QUIC connection is dropped and re-established, a node might miss slot updates. 

We cover all three cases in the same way. The networking layer internally keeps a \emph{connection ID} counter that is incremented with every new peer connection that is being established.
The networking layer uses the connection ID to ensure the connection is the same as before. It checks it every few seconds, and if the values are different, it triggers a \emph{retransmission} process: the entire content of the slot table is sent at once to the other side. Note that only the adverts are retransmitted, not entire messages, making the process fast. The corresponding peer will also notice the change in the connection ID and will trigger the same process in the opposite direction. Note that this means that aborted messages are not retransmitted.

\subsection{Bounding the Pools}
\label{sec:design:pool-bounds}
The client has the responsibility to bound the size of the validated message pool. But it is the networking layer's responsibility to bound the size of the unvalidated message pool, without relying on the client to evaluate the validity of messages in a timely manner. However, malicious peers may send as many messages as they want by sending slot updates that signal the replacement of content in their slot table (or pretending to do so).
Thus, nodes track their receive-side slot tables, and whenever a message is no longer advertised by any peer, it is automatically deleted from the unvalidated message pool. This still respects the guarantee \ref{guarantee:delivery}, and thus sufficient for the clients we consider (that posses the property~\ref{prop:expiry}). The size of the unvalidated message pool is therefore bounded to $O(C \cdot n \cdot \texttt{max\_message\_size})$ bytes in the worst case, since each of the $n$ sender nodes can advertise at most $C$ messages at any point in time. Together with the bounded size of the slot table, which gives us bounded internal data structures, this achieves the guarantee \ref{guarantee:bounded-ds}. This holds for recovering nodes as well. We note that the bound on the size of the unvalidated pool can be made tighter by exploiting some client-specific knowledge and the fact that honest peers generally relay the same messages, but we omit the details in this paper.


\section{Implementation}
\label{sec:impl}
The \ic system has been fully implemented, deployed in production, and has been running with minimal issues for nearly three years. The networking layer, which is the focus of this paper, has been in use since September 2023. The source code is available on \url{https://github.com/dfinity/ic}.

\begin{figure}
    \centering
    \includegraphics[width=\columnwidth]{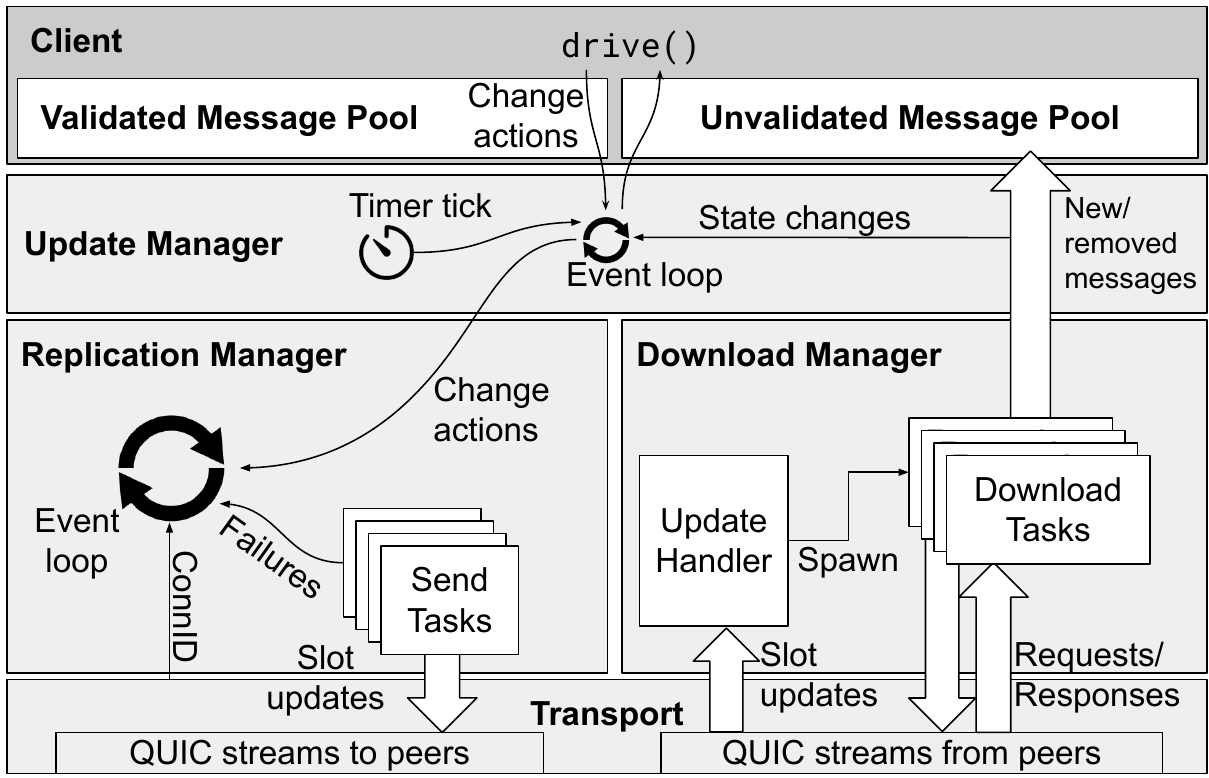}
    \vspace{-.6cm}
    \caption{Main components of the implementation.}
    \vspace{-.4cm}
    \label{fig:impl}
\end{figure}

Figure~\ref{fig:impl} depicts the main components of the implementation and their interactions. These components include the \textit{transport} library, which forms the core of the networking layer, the \textit{Replication Manager} (responsible for the send-side logic), and the \textit{Download Manager} (handling the receive-side logic). Additionally, the \textit{Update Manager} manages communication with message pools and triggers client events.

The \ic implementation is written in Rust. The networking layer is designed to be fully asynchronous and highly concurrent, with careful attention given to minimizing contention.  This is achieved through the use of the \texttt{tokio} asynchronous runtime library \cite{tokio}, which provides efficient concurrency in user-space with \emph{tasks} (lightweight threads). The transport component interfaces with the \texttt{quinn} library \cite{quinn} that implements IETF RFC 9000 \cite{quic} for QUIC. We assume that the CPU scheduler, the \texttt{tokio} scheduler, and the synchronization primitives used by \texttt{quinn} and the OS are fair, i.e., they do not starve tasks by never scheduling them.

\subsection{Transport}
\label{sec:impl:transport}
The \textbf{Transport} component is a key RPC-like library that enables multitenancy for the \ic consensus protocols on a single node. It abstracts reliable delivery and multiplexing for the entire networking layer and is implemented using the QUIC protocol.

We evaluated alternative implementations using TCP and UDP. While TCP may achieve better throughput than QUIC and be less susceptible to QoS-related packet drops over WANs, its single-stream nature can lead to head-of-line blocking for clients sharing the same connection.  Using multiple TCP connections or libraries like Yamux  \cite{yamux} could mitigate this, but these solutions increase complexity by requiring features already available in QUIC.

Although UDP performs well under optimal network conditions, it requires application-layer implementations for acknowledgments and lacks built-in flow and congestion control mechanisms, making it less ideal for our purposes.

\subsection{Replication Manager}
The  Replication Manager provides the nonblocking IO \textit{broadcast} and \textit{abort} operations. It runs a main thread with an event-loop, and a set of \emph{send tasks}. Each send task ensures the reliable delivery of the most recent content from a slot (or a set of slots) to one peer.

The main event-loop handles the following triggers:
\begin{itemize}
    \item \textbf{New Change Actions from a Client}:  These actions indicate modifications to the validated pool. The send-side slot table and send tasks are updated accordingly. Specifically, active tasks for deleted messages are terminated, and new tasks are spawned for newly added messages.
    \item \textbf{Send Task Failure}: Failures occur when a connection breaks or when the receiving application explicitly rejects an incoming message. The system retries delivery with an exponentially increasing random delay.
    \item \textbf{Change in Peer’s Connection ID}: This signal indicates that a peer’s connection has been dropped and re-established. Consequently, all associated send tasks are terminated, and new tasks are created for slots with valid content (as described in Section~\ref{sec:design:recovery}, this implements the retransmission process).

\end{itemize}

\subsection{Download Manager}
The \textbf{Download Manager} tracks incoming slot updates from peers and manages the download of advertised messages accepted by the bouncer function. The \emph{update handler} is triggered whenever a new slot update is received from a peer (multiple updates can be processed concurrently).  If slot content refers to a new message then a single download task is spawned. If multiple peers send slot content that refers to the same message, only one download task will be running, attempting, if needed, to download the message from any of the peers (specifically, from the first peer who advertised it, and if this fails or times out, from other peers advertising the same message). The download manager implements the slot tables on the receive side without storing the entire slot tables of all peers explicitly in memory. Instead, the slot table entry for each peer contains only a reference to the corresponding download task (which may correspond to multiple slot table entries from multiple peers). 

The download task for a message is alive (but perhaps idle) as long as a corresponding slot table entry exists at any of the peers, even after the message has been downloaded. This helps track peers broadcasting the same message. It also allows the removal of aborted messages from the unvalidated pool, as explained in Section~\ref{sec:design:pool-bounds}. Specifically, when a download task is terminated, the corresponding message is removed from the unvalidated pool (if such a message still exists).

When a peer requests a message download, a separate task, spawned by the transport library, retrieves the content from the validated pool and sends it to the requester.


\subsection{Update Manager}
\label{sec:impl:artifact-manager}
The \textbf{Update Manager} coordinates communication between a client, including its associated message pools, and the abortable broadcast implementation. It operates in a dedicated thread, running an event loop that processes two main types of events (1) insertion or removal of messages from the unvalidated pool and (2) timer ticks. Timer ticks occur every 200 milliseconds to ensure client progress, even in the absence of new messages from peers.
Each client operates as a state machine, with its state defined by the contents of its message pools. The transition function, \verb|drive()|, is triggered by the Update Manager for each event. When called, the client generates a set of change actions to update the validated pool. These changes are then passed to the Replication Manager for further processing, as previously described.

\section{Analysis}
\label{sec:analysis}

In this section, we give a high-level rationale on the security and correctness of the implementation.
We explain why it satisfies the guarantees~\ref{guarantee:delivery}-\ref{guarantee:partial-sync}. 

\textbf{Abortable guaranteed delivery}
For \ref{guarantee:delivery} we must demonstrate that a malicious peer cannot prevent $R$ from obtaining $m$ from $S$ eventually. The update handler interleaves updates from different peers, by imposing fair scheduling and a bounded number of concurrent streams for each peer, such that updates from honest peers will be received and processed even if a malicious peer sends a large number of messages or adverts. By advertising the same message $m$ as $S$, an attacker may delay the reception of $m$ at $R$ by not replying to a request, but eventually $R$'s download manager will request $m$ from $S$. While $S$ may be replying to an undue load of requests from malicious nodes, due to fair scheduling and the bounded number of concurrent streams for each peer, $S$ will eventually respond to requests from $R$. The recovery procedure ensures that nodes are able to re-transmit slot table updates and messages, thus \ref{guarantee:delivery} holds even when nodes may crash and recover, as long as both $R$ and $S$ are up and running for long enough.

\textbf{Integrity} The integrity property~\ref{guarantee:integrity} trivially follows since we use QUIC with TLS enabled.

\textbf{Bounded Datastructures}
We now argue that \ref{guarantee:bounded-ds} holds. 
The number of send tasks running is bounded by $(n-1)C$ and per connection a constant amount of information is stored. Analogously, on the receive side, the number of concurrent updates being handled is bounded by $(n-1) C$ and the slot table and unvalidated pool are bounded as discussed in Section~\ref{sec:design:pool-bounds}.

\textbf{Bounded-Time Delivery} We show~\ref{guarantee:partial-sync} by following the path of a message $m$ added by $S$ to the validated pool. As stated in~\ref{guarantee:partial-sync}, we assume that neither $S$ nor $R$ crash for some bounded period, and that $R$'s bouncer function accepts $m$. Since the send side table is bounded, fair scheduling of send tasks at $S$ guarantees that $S$ will send an advert for $m$ in bounded time (i.e., that it will enter the send-side QUIC buffer), regardless of the churn in other slots. Next, the "network behaving correctly" assumption in \ref{guarantee:partial-sync} implies that QUIC delivers the advert for $m$ to $R$ in bounded time. Fair scheduling of processing incoming messages from different peers by $R$ ensures that there is a bound on how much a malicious peer can delay $R$'s processing of an incoming slot update by $S$. When using adverts and separate message downloads, download timeouts and a round-robin selection of peers ensure that $R$ will --- within bounded time --- try to download the message from some correct peer $S'$ (who may be $S$), and not from a malicious peer. Since there is a bound on the number of concurrent streams for each peer at $S'$ and fair scheduling between them, since the messages are bounded in size and since we assume good network weather, $m$ will be received by $R$ in a bounded amount of time.

\section{Evaluation}
\label{sec:eval}
We evaluated our implementation by deploying it both to AWS, for running specific experiments, and to the \ic mainnet, to see how it works on a large scale of multiple peer groups running over a total of hundreds of autonomous nodes.

\begin{figure}[t]
    \centering
    \includegraphics[width=0.42\textwidth]{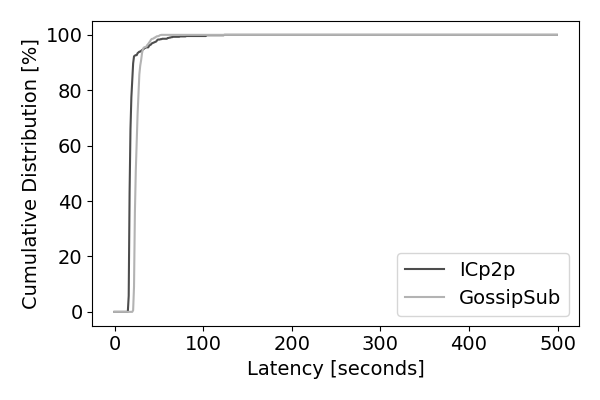}
    \caption{Latency comparison in a 52-node peer group.}
    \label{fig:52nodes:latency}
\end{figure}

We deployed a sanitized version of the networking layer, detached from other components of the \ic software stack (e.g., without the consensus protocol or any other client). Instead, in this version, a mock client automatically generates messages of a given size, at a given rate, places them in the validated message pool, and asks the networking layer to push all of them regardless of their size.

Since the \ic's mainnet is governed by decentralized voting, there is no simple way to deploy experimental code to it, and thus we used AWS for the experiments that are described in this section. 
We deployed the code to AWS \texttt{m7i.2xlarge} instances, with up to $12.5Gbps$ links.
We used nodes deployed in the same region, and applied traffic control (\texttt{tc}) rules to add $50ms$ latency on each link ($100ms$ RTT) to simulate a more geographically-dispersed environment similar to the latencies observed in practice in the \ic network and other geographically-dispersed BFT-based networks.

We compare our implementation for the networking layer, which we denote here as \emph{\icpp}, with a version that, instead of our networking layer code, uses \libpp's \gossipsub as the networking layer (with default parameters, using QUIC). We consider \gossipsub to be the state-of-the-art open-source peer-to-peer library (and the only one available and ready to use), and so we compare our implementation to it. 

\subsection{Performance Benchmark}
The two main metrics we are interested in are \emph{goodput} and \emph{latency}. We define goodput as the rate (in bytes/s) of unique message sending and receiving. Any overheads, such as adverts, messages that are relayed and therefore received more than once, and other protocol messages, do not count towards the goodput.
\icpp and \gossipsub achieve very similar goodput figures. Unsurprisingly, the two libraries are mainly limited by the implementation of the QUIC library and by the link bandwidth limits. Specifically, in a node group of size 52 nodes, with a message size of $5MB$ and a rate of 3 messages per second (each node sends and receives $3 \cdot 51$ messages per second), both get close to line rate ($12.5Gbps$ in our setup).

Figure~\ref{fig:52nodes:latency} shows a latency comparison between \icpp and \gossipsub, when running a peer group of 52 nodes, with a message size of $2MB$ and a send rate of approximately 4 messages per second (hence total bidirectional throughput of about $6.4Gbps$). The experiment runs for 15 minutes.
The latency in this experiment is the time it takes for a message from when it is placed in the validated pool of the sender until it gets to the unvalidated pool of the receiver. We use machines with synchronized clocks on AWS. The mock client adds a timestamp when placing a message in the validated pool, and it reads it on the receiver side when the message is placed in the unvalidated pool. The latency distribution is comparable, and this is consistent in multiple runs on different sizes of peer groups and with varying sizes of messages.

To measure potential network-layer message loss we ran the experiment with increasing rates of up to 20 messages per second from each peer, in a 31-nodes peer group, and a message size of $1MB$ (that is, up to $4.8Gbps$ send rate, up to $9.6Gbps$ send+receive per node, below link limit),
for 10 minutes, and then stopped message production. No broadcast abortions are executed. Our implementation delivers all messages to all peers as expected. \gossipsub shows a small percentage of $0.026\%$ message loss.

\begin{figure}[t]
    \centering
    \vspace{-.4cm}
    \includegraphics[width=0.47\textwidth]{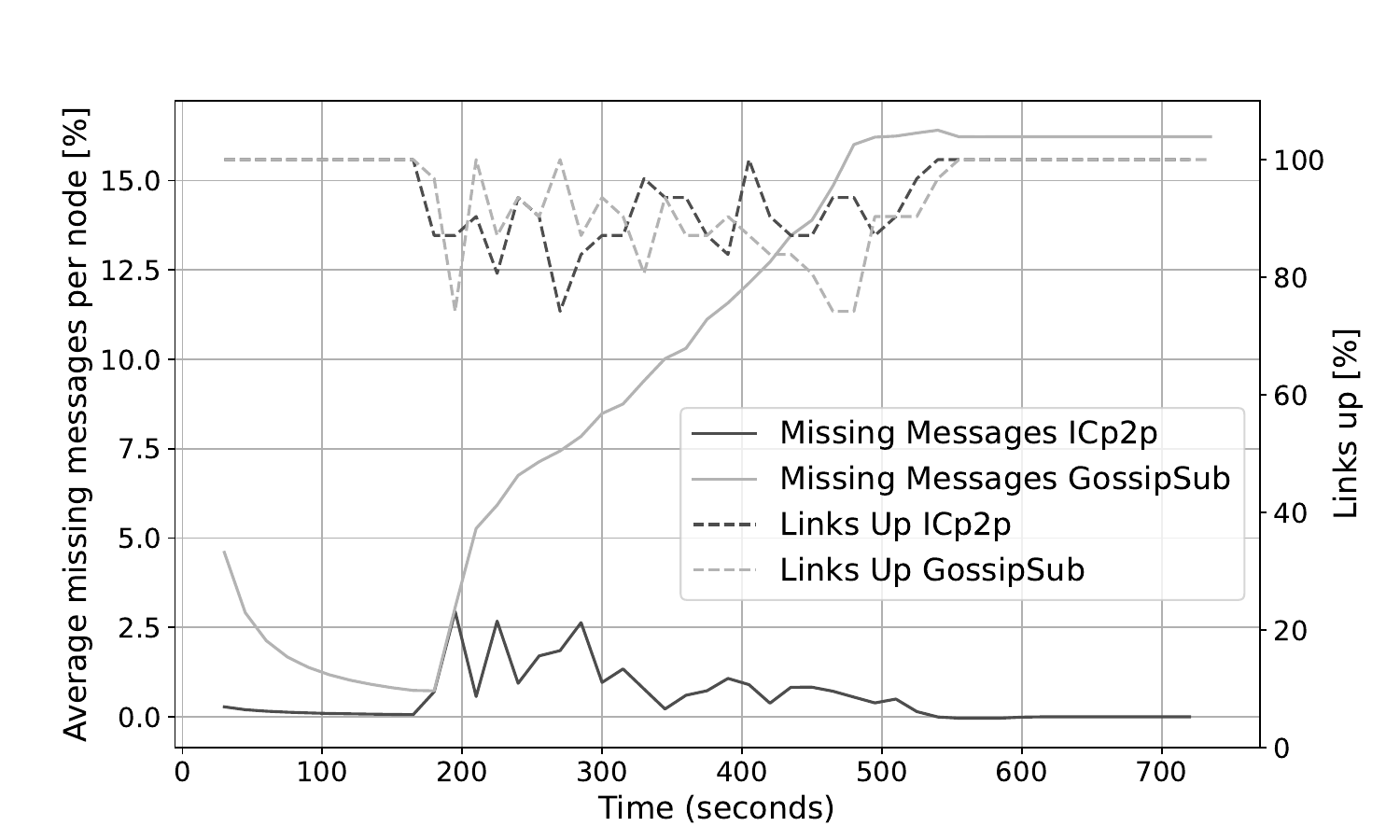}
    \caption{Message loss with network connectivity issues.}
    \label{fig:failing-links}
\end{figure}

\subsection{Operation With Faults}

We conduct experiments to demonstrate that our implementation avoids the issues outlined in the introduction. Specifically, we show that our non-blocking interface prevents message loss and maintains bounded memory usage.

\paragraph{Bad Links}
We now turn to examining how \icpp operates when faults occur. First, we look at a situation where nodes intermittently lose their network connectivity, uniformly at random, to understand how this affects message delivery at the client level. We ran an experiment with \icpp and \gossipsub on a 31 node network with each node sending one 200kb message every second. Between seconds 180--520 of the experiment, in every 30-second interval, each node has a $20\%$ chance of its connection to fail, and failures last 20 seconds. Figure~\ref{fig:failing-links} shows the number of messages that are missing on the receive side (averaged over all nodes), as a function of the time in the experiment. \icpp guaranteed delivery ensures that all messages are eventually delivered, so that the client does not miss any message it still needs, while \gossipsub without delivery guarantees simply loses more than 15\% of the messages indefinitely.

\begin{figure*}
    \centering
    \begin{subfigure}[b]{0.42\textwidth}
        \centering
        \includegraphics[width=\textwidth]{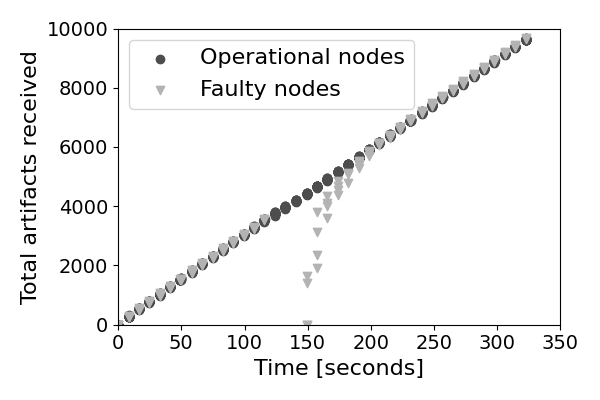}
        \vspace{-.5cm}
        \caption{\small{\icpp}}
        \label{fig:catchup:ours}
    \end{subfigure}
    \hfill
    \begin{subfigure}[b]{0.42\textwidth}
        \centering
        \includegraphics[width=\textwidth]{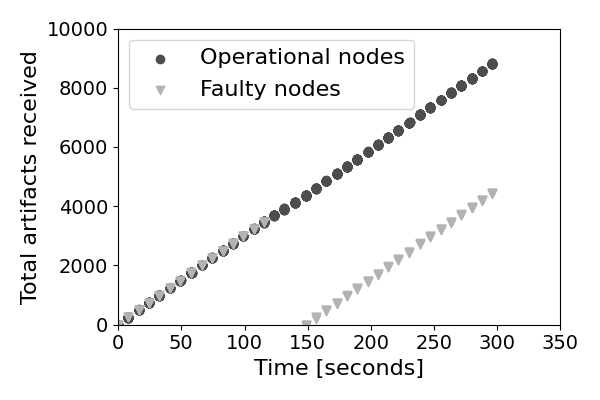}
        \vspace{-.5cm}
        \caption{\small{\gossipsub}}
        \label{fig:catchup:gossipsub}
    \end{subfigure}
    \vspace{-.2cm}
    \caption{Nodes catching up with missing messages after faults.}
    \vspace{-.4cm}
    \label{fig:catchup}
\end{figure*}

\paragraph{Node Crashes}
Figure~\ref{fig:catchup} shows how nodes catch up with missed messages after a crash or a period with connectivity problems. 
In this experiment, 4 nodes out of 31 were taken down for 30 seconds. With \icpp, these nodes quickly catch up after recovering, while with \gossipsub, as expected, they never catch up on the missed messages, hence additional means for retransmission must be implemented on top of it.

\begin{figure}[t]
    \centering
    \vspace{-.3cm}
    \includegraphics[width=0.48\textwidth]{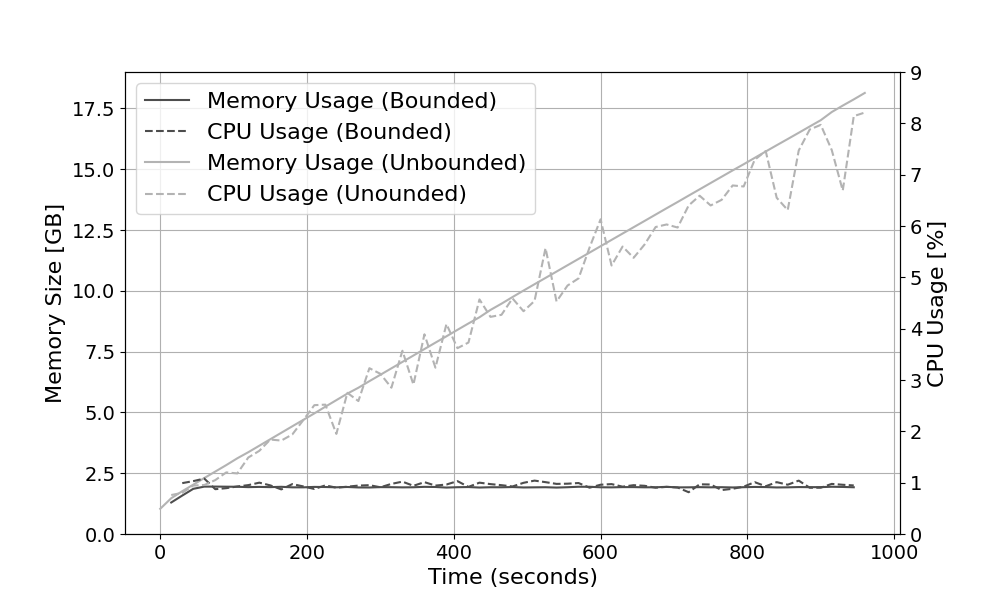}
    \vspace{-.6cm}
    \caption{Memory and CPU usage of nodes under attack.}
    \label{fig:memory}\vspace{-.3cm}
\end{figure}

\paragraph{Denial-of-Service Attack Scenario}
As discussed earlier in the paper, a malicious node may try to flood its peers with fake messages to fill up their data structures, slow them down, and potentially get their corresponding process beyond its memory bounds. This is the reason we argue for the need for the abort operation. Figure~\ref{fig:memory} shows the memory consumption and CPU usage of the \ic process on two victim nodes, as a function of the time. One victim node has a bound of $50K$ entries on its slot table, while the other one has no bound at all. During this time, two other nodes attack the victim nodes by sending each of them $1K$ new messages per second. The victim node with a bounded slot table maintains a steady memory usage a bit below 2GB and CPU usage around $1\%$ (for the entire process, not just for \icpp), while the other victim node, with no bounds on the slot table, keeps growing its memory and CPU usage.

\subsection{\ic Mainnet Deployment Observations}
\label{sec:eval:mainnet:slottable}
To complete the picture about the environment for which we designed the system,  we provide data collected from the mainnet deployment of the \ic, where user applications are deployed and run. We believe this data can be useful to complete the picture about the environment in which such systems operate.
The \ic is built of nodes running in independent DCs worldwide, connected over the public IPv6 Internet without dedicated links. Peer groups are geographically dispersed for maximal decentralization. The median inter-continental cross-DC RTT exposed by metrics of the \texttt{quinn} QUIC library observed over a span of 24 hours varies between at least $100ms$ RTT and up to $280ms$.  in some cases. Thus, there is a limit on how much we can optimize the protocol, and how fast messages can be delivered. Accordingly, for a typical peer group of highly dispersed nodes, the median RPC message download duration for the consensus client is 101ms, with 6\% of the messages taking 300-350ms.   
The observed number of slot table entries for such a peer group is below 5000, with around 20000 active download tasks. The mean slot table update rate varies between 110 to 325 per second for different peer groups. 

\section{Related Work}
\label{sec:related}
While there is an extensive line of research on Byzantine fault-tolerant distributed systems to be used in practice, from the well-known PBFT~\cite{pbft} to a variety of approaches in blockchain networks e.g.,~\cite{hotstuff,redbelly,bano2019sok,garay2020sok}, the vast majority of these works ignore the problems presented in this paper, resulting in protocols that are correct on paper, but are very hard to implement correctly without introducing vulnerabilities on the networking layer. The main reason is that implementing them according to their description leads to either potentially unbounded memory use, exposure to DoS attacks that can slow down the network, or loss of messages and thus liveness guarantees. For blockchain networks that are supposed to be autonomous and run forever, even a small probability of losing liveness means that there is a chance that the network will get stuck permanently, with no automated recovery option. 

Notably, Bitcoin has de-facto given up on the bounded memory requirement, with (full) nodes having to store an ever growing set of data necessary to validate block payloads. Therefore it can take days to sync when a new node joins~\cite{bitcoin_full_node}. This design choice is only possible due to the low throughput of this network, with Bitcoin handling around 7 tx/s. While Ethereum's throughput of around 12 tx/s is not much higher, its nodes can prune old data~\cite{ethereum_weak_subjectivity}.
Ethereum as well as other blockchain networks like IPFS~\cite{ipfs-sigcomm} rely on a 
\libpp (and specifically its gossip library \gossipsub), an example of an open-source, fully operational networking layer designed specifically for blockchain networks. However, as shown in Section~\ref{sec:eval}, \gossipsub does not guarantee the delivery of messages \cite{gossipsub-params}, since it drops messages in case of node crashes or if peers are too slow to receive them.

Aguilera et al.\cite{aguilera_ubft_2023} implement BFT consensus with bounded memory in a data center setting under the $f < n/2$ assumption, using certain types of disaggregated memory as a trusted computing base. In the process, they define a \emph{tail broadcast} primitive. The primitive is similar to abortable broadcast: it is also a best-effort broadcast with a weaker validity property, and it can be implemented with bounded memory. A significant difference, however, is that they do not change the broadcast interface, only the guarantees: they only guarantee delivery of the last $t$ broadcast messages, for some constant parameter $t$. This makes their primitive less versatile than abortable broadcast. For example, it suffices for their consensus algorithm, but not for the \ic algorithms. Conversely, abortably broadcast can trivially implement tail broadcast (and thus their consensus algorithm) without overhead: set the capacity $C$ to $t$, and simply abort everything but the last $t$ messages. Implementing abortable broadcast using tail broadcast as a black box is conceptually also possible:  set $t = C$, store all the active messages and resend \emph{all} of them once any message is aborted. But this introduces an impractical overhead. Design-wise, their system is focused on the local data-center setting and small messages (their evaluation covers cases up to 2kb), and assumes the existence of disaggregated memory and RDMA communication, with tail broadcast using RDMA only. Implementing tail broadcast over WAN is not discussed in the paper. A straightforward implementation would use UDP, with all the drawbacks mentioned in Section~\ref{sec:impl:transport}. Thus, their proposed design and implementation is unsuitable for BFT protocols designed for WAN, such as public blockchains.
\yotam{I changed the end to be more concrete.}

Castro and Liskov \cite{castro01practical,pbft}, in their seminal work on practical Byzantine fault-tolerance, address networking issues to some extent, and rightfully claim that simply using any sort of reliable transport would mean unbounded buffering if we want to guarantee correctness (see Section 5.2 in \cite{castro01practical}). Instead, similar to our naive protocol from Section~\ref{sec:ab:algo} they use unreliable message transport and periodic retransmission of lost messages. However, in their approach, the transmission is always requested by the receiver. This approach, however, has several problems when implemented in practice: first, for protocols with messages bigger than MTU size, it requires correct framing and defragmentation of messages, which in turn requires a reordering buffer. Second, their retransmission mechanism is highly specific to the PBFT algorithm, whereas the networking layer proposed in this paper is client-agnostic.

Shahmirzadi et al.~\cite{shahmirzadi_relaxed_2009} show how to implement a variant of consensus using bounded memory, but do not provide a modular broadcast primitive as such.

\section{Future Work}
\label{sec:future}
The \ic protocol stack and its networking layer are continuously evolving, with new protocol versions being adopted roughly weekly. We invite readers to collaborate with us on open problems.
For example, we strive to increase the scalability of the protocol with respect to peer group size while maintaining high throughput, and to grow the number of groups while keeping cross-group communication (not discussed in this paper for brevity) efficient.
Furthermore, protection against volumetric DoS attacks, e.g., with scrubbing adapted to decentralized environments where no single vendor can be trusted~\cite{ddos-blockchain} is of high interest.

Instead of focusing on implementing a practical variant of the best-effort broadcast primitive, one could also consider abortable versions of other broadcast variants. For example, we believe that the approach in Section~\ref{sec:design} can be modified to provide an abortable (and BFT) version of FIFO broadcast~\cite{cachin2011introduction}, which provides ordering guarantees. Clients may be rewritten to take advantage of this ordering to eliminate the unvalidated pool, and thus further lower the memory usage on the receive side to a constant amount per peer. Similarly, abortable versions of consistent and reliable broadcast could also be of independent interest.

Other research topics include node behavior monitoring, load balancing and caching in the blockchain settings, as well as decentralizing service addressing and DNS.

\section{Conclusions}
\label{sec:conclusions}

In this paper, we proposed a new generic broadcast primitive that slightly weakens the requirement of guaranteed delivery, and in turn allows simple and efficient implementation of the primitive using only bounded amounts of memory. Furthermore, we evaluated it both on synthetic benchmarks and in the production environment of the \ic's globally distributed blockchain protocol, and found it compared favorably with the state of the art.
However, the importance of keeping memory bounded is obviously relevant for any BFT protocol, not just the blockchain ones. We believe that our work provides a generic and reusable building block for achieving this requirement.



\label{docuement-end}

\bibliographystyle{plain}
\bibliography{refs}

\begin{thebibliography}{10}

\bibitem{aguilera_ubft_2023}
Marcos~K. Aguilera, Naama Ben-David, Rachid Guerraoui, Antoine Murat, Athanasios Xygkis, and Igor Zablotchi.
\newblock {uBFT}: {Microsecond}-{Scale} {BFT} using {Disaggregated} {Memory}.
\newblock In {\em Proceedings of the 28th {ACM} {International} {Conference} on {Architectural} {Support} for {Programming} {Languages} and {Operating} {Systems}, {Volume} 2}, pages 862--877, Vancouver BC Canada, January 2023. ACM.

\bibitem{aumasson_survey_2020}
Jean-Philippe Aumasson, Adrian Hamelink, and Omer Shlomovits.
\newblock A survey of {ECDSA} threshold signing.
\newblock {\em Cryptology ePrint Archive}, 2020.

\bibitem{avarikioti_fnf-bft_2023}
Zeta Avarikioti, Lioba Heimbach, Roland Schmid, Laurent Vanbever, Roger Wattenhofer, and Patrick Wintermeyer.
\newblock {FnF}-{BFT}: {A} {BFT} {Protocol} with {Provable} {Performance} {Under} {Attack}.
\newblock volume 13892, pages 165--198, Cham, 2023. Springer Nature Switzerland.
\newblock Series Title: Lecture Notes in Computer Science.

\bibitem{bano2019sok}
Shehar Bano, Alberto Sonnino, Mustafa Al-Bassam, Sarah Azouvi, Patrick McCorry, Sarah Meiklejohn, and George Danezis.
\newblock Sok: Consensus in the age of blockchains.
\newblock In {\em Proceedings of the 1st ACM Conference on Advances in Financial Technologies}, pages 183--198, 2019.

\bibitem{bitcoin_full_node}
{Bitcoin Project}.
\newblock Running {A} {Full} {Node} - {Bitcoin}, Accessed 2024.
\newblock \url{https://bitcoin.org/en/full-node}.

\bibitem{buchman_tendermint_2016}
Ethan Buchman.
\newblock {\em Tendermint: {Byzantine} fault tolerance in the age of blockchains}.
\newblock {PhD} {Thesis}, University of Guelph, 2016.

\bibitem{buterin_casper_2019}
Vitalik Buterin and Virgil Griffith.
\newblock Casper the {Friendly} {Finality} {Gadget}, January 2019.
\newblock arXiv:1710.09437 [cs].

\bibitem{cachin2011introduction}
Christian Cachin, Rachid Guerraoui, and Lu{\'\i}s Rodrigues.
\newblock {\em Introduction to reliable and secure distributed programming}.
\newblock Springer Science \& Business Media, 2011.

\bibitem{camenisch2022internet}
Jan Camenisch, Manu Drijvers, Timo Hanke, Yvonne-Anne Pignolet, Victor Shoup, and Dominic Williams.
\newblock Internet computer consensus.
\newblock In {\em Proceedings of the 2022 ACM Symposium on Principles of Distributed Computing}, pages 81--91, 2022.

\bibitem{castro01practical}
Miguel Castro.
\newblock {\em Practical Byzantine Fault Tolerance}.
\newblock {Ph.D.}, MIT, January 2001.
\newblock Also as Technical Report MIT-LCS-TR-817.

\bibitem{pbft}
Miguel Castro and Barbara Liskov.
\newblock Practical byzantine fault tolerance.
\newblock In {\em 3rd Symposium on Operating Systems Design and Implementation (OSDI 99)}, New Orleans, LA, February 1999. USENIX Association.

\bibitem{ddos-blockchain}
Rajasekhar Chaganti, Bharat Bhushan, and Vinayakumar Ravi.
\newblock The role of blockchain in ddos attacks mitigation: techniques, open challenges and future directions.
\newblock {\em CoRR}, abs/2202.03617, 2022.

\bibitem{redbelly}
Tyler Crain, Christopher Natoli, and Vincent Gramoli.
\newblock Red belly: {A} secure, fair and scalable open blockchain.
\newblock In {\em 42nd {IEEE} Symposium on Security and Privacy, {SP} 2021, San Francisco, CA, USA, 24-27 May 2021}, pages 466--483. {IEEE}, 2021.

\bibitem{das_practical_2022}
Sourav Das, Thomas Yurek, Zhuolun Xiang, Andrew Miller, Lefteris Kokoris-Kogias, and Ling Ren.
\newblock Practical asynchronous distributed key generation.
\newblock In {\em 2022 {IEEE} {Symposium} on {Security} and {Privacy} ({SP})}, pages 2518--2534. IEEE, 2022.

\bibitem{ethereum_weak_subjectivity}
{Ethereum Foundation}.
\newblock Weak subjectivity, Accessed 2024.
\newblock \url{https://ethereum.org/developers/docs/consensus-mechanisms/pos/weak-subjectivity}.

\bibitem{garay2020sok}
Juan Garay and Aggelos Kiayias.
\newblock Sok: A consensus taxonomy in the blockchain era.
\newblock In {\em Cryptographers’ track at the RSA conference}, pages 284--318. Springer, 2020.

\bibitem{tecdsa}
Jens Groth and Victor Shoup.
\newblock Design and analysis of a distributed ecdsa signing service.
\newblock Cryptology ePrint Archive, Paper 2022/506, 2022.
\newblock \url{https://eprint.iacr.org/2022/506}.

\bibitem{dkg}
Jens Groth and Victor Shoup.
\newblock Fast batched asynchronous distributed key generation.
\newblock In {\em Advances in Cryptology - {EUROCRYPT} 2024 - 43rd Annual International Conference on the Theory and Applications of Cryptographic Techniques, Zurich, Switzerland, May 26-30, 2024, Proceedings, Part {V}}, volume 14655 of {\em Lecture Notes in Computer Science}, pages 370--400. Springer, 2024.

\bibitem{yamux}
{HashiCorp}.
\newblock Yamux, Accessed 2024.
\newblock \url{https://github.com/hashicorp/yamux}.

\bibitem{quic}
Jana Iyengar and Martin Thomson.
\newblock {QUIC: A UDP-Based Multiplexed and Secure Transport}.
\newblock RFC 9000, May 2021.

\bibitem{lamport_byzantine_1982}
Leslie Lamport, Robert Shostak, and Marshall Pease.
\newblock The {Byzantine} {Generals} {Problem}.
\newblock {\em ACM Transactions on Programming Languages and Systems}, 4(3), 1982.

\bibitem{tokio}
Carl Lerche, Accessed 2024.
\newblock tokio \url{https://tokio.rs/}.

\bibitem{lu_honeybadgermpc_2019}
Donghang Lu, Thomas Yurek, Samarth Kulshreshtha, Rahul Govind, Aniket Kate, and Andrew Miller.
\newblock Honeybadgermpc and asynchromix: {Practical} asynchronous mpc and its application to anonymous communication.
\newblock In {\em Proceedings of the 2019 {ACM} {SIGSAC} {Conference} on {Computer} and {Communications} {Security}}, pages 887--903, 2019.

\bibitem{quinn}
Dirkjan Ochtman, Benjamin Saunders, and Jean-Christophe Begue, Accessed 2024.
\newblock quinn \url{https://github.com/quinn-rs/quinn}.

\bibitem{libp2p}
{Protocol Labs}, Accessed 2024.
\newblock libp2p \url{https://libp2p.io/}.

\bibitem{gossipsub-params}
{Protocol Labs}, Accessed 2024.
\newblock gossipsub v1.1: Security extensions to improve on attack resilience and bootstrapping \url{https://github.com/libp2p/specs/blob/master/pubsub/gossipsub/gossipsub-v1.1.md}.

\bibitem{shahmirzadi_relaxed_2009}
Omid Shahmirzadi, Sergio Mena, and Andre Schiper.
\newblock Relaxed {Atomic} {Broadcast}: {State}-{Machine} {Replication} {Using} {Bounded} {Memory}.
\newblock In {\em 2009 28th {IEEE} {International} {Symposium} on {Reliable} {Distributed} {Systems}}, pages 3--11, Niagara Falls, New York, USA, September 2009. IEEE.

\bibitem{stathakopoulou_mir-bft_2019}
Chrysoula Stathakopoulou, Tudor David, and Marko Vukolic.
\newblock Mir-bft: {High}-throughput bft for blockchains.
\newblock {\em arXiv preprint arXiv:1906.05552}, 92, 2019.

\bibitem{ipfs-sigcomm}
Dennis Trautwein, Aravindh Raman, Gareth Tyson, Ignacio Castro, Will Scott, Moritz Schubotz, Bela Gipp, and Yiannis Psaras.
\newblock Design and evaluation of ipfs: a storage layer for the decentralized web.
\newblock In {\em Proceedings of the ACM SIGCOMM 2022 Conference}, SIGCOMM '22, page 739–752, New York, NY, USA, 2022. Association for Computing Machinery.

\bibitem{gossipsub}
Dimitris Vyzovitis, Yusef Napora, Dirk McCormick, David Dias, and Yiannis Psaras.
\newblock Gossipsub: Attack-resilient message propagation in the filecoin and {ETH2.0} networks.
\newblock {\em CoRR}, abs/2007.02754, 2020.

\bibitem{hotstuff}
Maofan Yin, Dahlia Malkhi, Michael~K. Reiter, Guy Golan{-}Gueta, and Ittai Abraham.
\newblock Hotstuff: {BFT} consensus with linearity and responsiveness.
\newblock In Peter Robinson and Faith Ellen, editors, {\em Proceedings of the 2019 {ACM} Symposium on Principles of Distributed Computing, {PODC} 2019, Toronto, ON, Canada, July 29 - August 2, 2019}, pages 347--356. {ACM}, 2019.

\end{thebibliography}

\label{bib-end}

\end{document}